\documentclass[aps,prb,twocolumn]{revtex4}
\usepackage{graphicx}
\usepackage{dcolumn}
\usepackage{amsmath}
\usepackage{amssymb}
\usepackage{amsmath,verbatim}
\usepackage{bm}

\def\rv{{\bf r}}
\def\jv{{\bf j}}

\def\zv{{\bf z}}
\def\uv{{\bf u}}

\begin{document}
\title{Lorentz shear modulus of a two-dimensional electron gas at high magnetic field} 
\author{I. V. Tokatly}
\affiliation{
Moscow Institute of Electronic Technology, Zelenograd, 124498 Russia}
\author{G. Vignale}
\affiliation{Department of Physics, University of Missouri-Columbia,
Columbia, Missouri 65211}
\date{\today}
\begin{abstract}
We show that the Lorentz shear modulus -- one of the three elastic moduli of a homogeneous electron gas in a magnetic field --  can be calculated exactly in the limit of high magnetic field (i.e. in the lowest Landau level).  Its value is $\pm \hbar n/4$, where $n$ is the two-dimensional electron density and the sign is determined by the orientation of the magnetic field. We use this result to refine our previous calculations of the dispersion of the collective modes of fractional quantum Hall liquids.
\end{abstract}
\maketitle
The collective dynamics of a two-dimensional electron gas at high magnetic fields continues to be a subject of intense interest and a source of surprises.  Witness  be the recent observation, by inelastic light scattering,  of a new collective mode in the fractional quantum Hall liquid at filling factor $\nu=1/3$.\cite{Hirjibehedin2005}  Theoretically, the interest arises from the fact that the electrons in the lowest Landau level (LLL) exhibit an unusual kind of collective behavior (for the most recent overview of the theory of fractional quantum Hall effect see, e.~g., Ref.~\onlinecite{JainBook}). In an ordinary two-dimensional quantum liquid collective modes arise either as hydrodynamic modes (sustained by frequent collisions in local quasi-equilibrium) or as collisionless modes sustained by strong self-consistent fields (in which case a generalized hydrodynamic description is possible). In both cases the frequency of the modes tend to zero at long wavelength, reflecting the underlying translational invariance of the system.  By contrast, in the incompressible fractional quantum Hall liquid the collective modes have a finite frequency in the long wavelength limit.  One mode  at the classical cyclotron frequency $\omega_c = \frac{eB}{m}$  ($-e$ and $m$ being the electron charge and mass -- we use SI units throughout)  is completely expected as a consequence of Kohn's theorem \cite{Kohn1961}. The surprise comes from the low-frequency mode, which in a fractional quantum Hall liquid (e.g. the $\nu=1/3$ state)   tends to a finite frequency $\Delta$ which scales as  $\frac{e^2}{\ell \epsilon_b}$ ($\ell=\sqrt{\frac{\hbar}{eB}}$ being the magnetic length).  The  gap has long been understood as a manifestation of the incompressibility of the electron liquid in the LLL~\cite{GirMacPla1986,ScaParJai2000}, which is also  responsible for the fractional quantum Hall effect \cite{Laughlin1983}.  (The gap closes in compressible states, e.g. the  $\nu=1/2$ state.)

The presence of the gap raises the question  whether a continuum mechanics formulation (hydrodynamics/elasticity)  for this type of modes is possible at all.  At first sight such a formulation can only produces gapless modes.  Yet, the absence of low-energy excitations or other single particle excitations overlapping the energy of the collective mode strongly suggests that a collective description of the dynamics should be possible.   Early attempts to formulate such a description relied on the ad-hoc introduction of a diverging bulk modulus \cite{ConVig1998}.    Recently, it has been found that the problem has a natural formulation as an elasticity theory in which the elastic constants are local in space, but strongly retarded in time \cite{TokatlyPRB2006a,TokatlyPRB2006b,TokVigPRL2007}.  Let us briefly review the essential points of this formulation.

First, the equation of motion for a Fourier component of the particle current density at frequency $\omega$, in the linear approximation and in the absence of external fields (other than a magnetic field $B$ in the negative $z$ direction) is
\begin{equation}\label{eom1}
-im\omega  \jv(\rv,\omega)-eB \jv(\rv,\omega) \times \hat \zv +  \nabla \cdot \stackrel{\leftrightarrow}{{\bf P}}(\rv,\omega)=0
\end {equation}
where $-e$ and $m$ are the electron charge and mass respectively,  $\stackrel{\leftrightarrow}{{\bf P}}(\rv,\omega)$ is the Fourier component of the stress tensor, and $\nabla \cdot$ denotes its divergence.  The stress tensor is defined as the expectation value of the stress tensor operator ${\bf \hat P}$, which can be directly derived  from the Heisenberg equation of motion for the current operator \cite{PufGil1968,TokatlyPRB2005a}.   

Second, the stress tensor is expressed (in the long-wavelength limit) as a  linear function of the current density
\begin{equation} \label{Pij}
P_{ij}(\rv,\omega) = -Q_{ijkl}(\omega)u_{kl}(\rv,\omega)
\end{equation}
(we sum over repeated indices)  where  
\begin{equation}
u_{kl}(\rv,\omega) = \frac{i}{2 n \omega} \left[\partial_k j_l(\rv,\omega) +  \partial_l j_k(\rv,\omega)\right],
\end{equation}
is the Fourier component of the {\it strain tensor} ($n$ being the homogeneous density of the electron gas),   and $Q_{ijkl}(\omega)$ is  the Fourier component of the homogeneous {\it tensor of elasticity} --  a fourth-rank tensor, which is symmetric with respect to any of the interchanges  $(ij) \leftrightarrow(kl)$, $i\leftrightarrow j$, or $k\leftrightarrow l$.   In two dimensions $Q_{ijkl}(\omega)$  has at most $6$ independent components. 
However, in a two-dimensional homogeneous electron liquid in the presence of a perpendicular magnetic field the six components of $Q_{ijkl}(\omega)$ can be expressed in terms of only three independent elastic moduli in the following manner:  $Q_{xx,xx}(\omega)=Q_{yy,yy}(\omega)=K(\omega)+\mu(\omega)$, $Q_{xy,xy}(\omega)=\mu(\omega)$, $Q_{xx,yy}(\omega)=K(\omega)-\mu(\omega)$, and $Q_{xx,xy}(\omega)=-Q_{yy,yx}(\omega)=i\omega \Lambda(\omega)$.  A compact representation is
\begin{eqnarray}\label{Qh1}
Q_{ijkl}(\omega) &=&K(\omega)\delta_{ij}\delta_{kl} +\mu(\omega)(\delta_{ik}\delta_{jl}+\delta_{jk}\delta_{il}-\delta_{ij}\delta_{kl})\nonumber \\ &+&i\omega \frac{\Lambda(\omega)}{2}(\varepsilon_{ik}\delta_{jl}+\varepsilon_{jk}\delta_{il}+
\varepsilon_{il}\delta_{jk}+\varepsilon_{jl}\delta_{ik}).\nonumber\\
\end{eqnarray}
where $K(\omega)$, $\mu(\omega)$ and $\Lambda(\omega)$ are the frequency-dependent bulk modulus, the shear modulus, and the Lorentz shear modulus respectively~\cite{note1}.   The first two moduli, $K$ and $\mu$, are familiar from conventional elasticity theory \cite{LandauVII:e}, and cannot be calculated without detailed microscopic input (notice however that the zero frequency limit of the shear modulus must be zero in a liquid state).   The Lorentz shear modulus $\Lambda$, introduced in Refs.~\onlinecite{TokatlyPRB2006a,TokatlyPRB2006b,TokVigPRL2007}, is a novel feature of the system in the presence of a magnetic field. Physically it is responsible for the force that squeezes together two oppositely directed streamlines. Formally it arises as follows:  First, rotational invariance about the $z$ axis mandates that the tensor of elasticity be invariant under the transformation $x \to y$, $y \to -x$:  this enforces the identity $Q_{xx,xy}(\omega)=-Q_{yy,yx}(\omega)$ at all frequencies.  Second, {\it at zero frequency} one has an additional reflection symmetry which allows us to interchange $x$ and $y$ without the minus sign: this implies $Q_{xx,xy}(0)=-Q_{yy,yx}(0)=0$.  In the absence of a magnetic field this symmetry persists even at finite frequency, but the magnetic field breaks it:  hence $\Lambda(\omega)$ has a finite non-zero value. 

The main purpose of this Communication is to show that the zero frequency limit of the Lorentz shear modulus $\Lambda_0  \equiv -\lim_{\omega \to 0} \Lambda(\omega)$  can be calculated exactly and has the universal value $\pm \hbar n/4$ ($+$ if the magnetic field is along the negative z axis, $-$ if it is along the positive $z$ axis) when all the electrons are in the lowest Landau level.  The reason is that $\Lambda_0$ can be expressed as a Berry curvature of the ground-state wave function with respect to a homogeneous change of the metrics, just as the Hall conductivity can be expressed as a Berry curvature with respect to a homogeneous vector potential~\cite{AvrSei1985,NiuThoWu1985}.  This analogy was first noted in a paper by Avron {\it et al.}\cite{AvrSeiZog1995}, entitled  ``The viscosity of the quantum Hall liquid", in which $\Lambda_0$ was calculated for a non interacting electron gas at integral filling factor, and improperly called  ``viscosity".  Because of the non-dissipative character of the dynamics we will continue to refer to $\Lambda_0$  as the ``Lorentz shear modulus".  Unlike Avron {\it et al.},  who considered only noninteracting systems at integral filling factors, we will calculate the Lorentz shear modulus for an interacting electron gas in a partially occupied lowest Landau level.   We will show that the calculation can be done exactly due to a special analyticity property which every  wave function in the LLL enjoys.   Armed with the exact Lorentz shear modulus we will then return to the equation of motion (\ref{eom1}) and show how the dispersion of the collective modes obtained in Ref.~\cite{TokVigPRL2007} by treating $\Lambda_0$  as a fit parameter, gets modified when the exact value of $\Lambda_0$ is used instead.

A very useful representation of the stress tensor can be given in terms of the derivative of  the Hamiltonian with respect to a metric tensor $g_{ij}(\rv)$ ($ds^2=g_{ij}dx^idx^j$)  which defines a non-euclidean geometry in the plane ($g_{ij}=\delta_{ij}$ is the Euclidean metrics) (see, e.~g., \cite{RogRap2002,TokatlyPRB2005a}).  This is the analogue of defining the current operator as the derivative of the Hamiltonian with respect to the vector potential \cite{ForRom2004}.   To this end, let us introduce the Hamiltonian of the spinless two-dimensional  electron gas in a metric field $g_{ij}(\rv)$:   
\begin{equation}
\label{H}
 H[{\bf g}]= \sum_n T_n  +\frac{1}{2} \sum_{n \neq n'} \frac{e^2}{\epsilon_b d(\rv_n,\rv_{n'})}
\end{equation}
where the sums run over particle indices ($n$ and $n'$), 
\begin{equation} \label{TN}
T_n =-\frac{1}{2m}g^{-\frac{1}{4}}(\rv_n) D_{n,i} \sqrt{g(\rv_n)}g^{ij}(\rv_n) D_{n,j}g^{-\frac{1}{4}}(\rv_n)
\end{equation}
is the kinetic energy of the $n$-th electron, $g^{ij}$ is the inverse of $g_{ij}$, 
\begin{equation}
D_{n,i} = \partial_{r_{ni}} + i e A_i (\rv_n)
\end{equation}
is the kinetic momentum operator,  ${\bf A}(\rv)$ is the vector potential, and $d[\rv,\rv']$ is the length of the geodesics connecting $\rv$ to $\rv'$ in the non-Euclidean plane.  We will work in the Landau gauge ${\bf A}(\rv) = (By,0)$ with periodic boundary conditions in the $x$ direction and open boundary conditions in the $y$ direction.  The linear size of the system in either direction is $L$, which we take as our unit of length, $L=1$.  We have also set $\hbar=1$.  In these units the operator ${\bf D}$ is conveniently expressed as
\begin{equation}
D_x=\partial_x+ 2 \pi i N_Ly,~~~~~D_y = \partial_y,
 \end{equation}
 where $N_L$ is the number of magnetic flux quanta ($h/e$) enclosed by the system.
 
 The stress tensor is defined as follows
 \begin{equation} \label{stresstensor}
 P_{ij}(\rv,t) = 2 \left\langle \frac{\delta H[{\bf g}]}{\delta g^{ij}(\rv)}\right\rangle
 \end{equation} where the average is taken in the time-dependent quantum state and at Euclidean metric.
 With this definition of the stress tensor, the equation of motion~(\ref{eom1}) is exact.  The basic task of the theory is to express the stress tensor as a (linear) functional of the current density  so that Eq.~(\ref{eom1}) can be closed.  This is a formidable task, of course, but we know that it can be carried out exactly in principle~\cite{Runge84}.

  As a first step in this direction we notice a small deviation from the Euclidean metrics, arising from the infinitesimal deformation $\rv \to \rv+\uv(\rv)$ ($\uv$ being the  integral in time of the velocity field $\jv/n$),  can be represented in the form $g^{ij}=\delta_{ij}-2 u_{ij}$ where $u_{ij}$ is the strain tensor.  Then, it is not difficult to show that $Q_{ijkl}(\omega)$ has the microscopic representation \cite{TokatlyPRB2005b}
\begin{equation}\label{Qh2}
Q_{ijkl}(\omega) = Q^\infty_{ijkl}+\langle\langle \hat P_{ij};\hat P_{kl}\rangle\rangle_\omega
\end{equation}
where the first term (purely real and independent of frequency) is the instantaneous derivative of $P_{ij}$ with respect to $g^{kl}$ and the second term is the Kubo formula for the stress-stress response function.

It is evident from Eq.~(\ref{Qh1}) that
\begin{equation}
\Lambda_0 = - \lim_{\omega \to 0} \Im m \frac{Q_{xxxy}(\omega)}{\omega}.
\end{equation}
Then, using Eq.~(\ref{Qh2}) in combination with the geometric definition of the stress tensor~(\ref{stresstensor}) we arrive, after well-known manipulations analogous to the ones that lead to the expression for the Hall conductivity \cite{AvrSei1985,NiuThoWu1985}, to the key result
\begin{equation}\label{key}
\Lambda_0 = 
8 \times \Im m\left. \left\langle \frac{\partial \psi_0}{\partial g^{xx}}\right \vert  \frac{\partial \psi_0}{\partial g^{xy}}\right\rangle,
\end{equation}
where $|\psi_0 \rangle$ is the ground-state of the Hamiltonian (\ref{H}) in the presence of a {\it homogeneous} metrics $g^{ij}$.  This is essentially the formula obtained in Ref.~\onlinecite{AvrSeiZog1995} for what they call ``viscosity".   

In order to proceed, it is very convenient to parametrize the metric tensor as follows
\begin{equation} \label{metric}
g_{ij}= \frac{J}{\tau_2}\left(\begin{array}{cc}  1 & \tau_1 \\ \tau_1 & |\tau|^2 \end{array}\right),~{\rm and}~g^{ij}= \frac{1}{J\tau_2}\left(\begin{array}{cc}  |\tau|^2 & -\tau_1 \\ -\tau_1 & 1 \end{array}\right),
\end{equation}
where $\tau =\tau_1+i \tau_2$ is a complex number defining the  length and orientation of the $y$ axis of the distorted plane, and $J$ is the Jacobian of the  coordinate transformation that is induced by deformation. The Euclidean metrics is recovered by setting $J=1$, $\tau_1 = 0$ and $\tau_2=1$.   
The derivatives with respect to $g^{ij}$ evaluated  at the Euclidean metrics are given by
\begin{eqnarray}\label{derivatives}
&&\frac{\partial}{\partial g^{xx}}=\frac{1}{2}\left(\frac{\partial}{\partial \tau_2}-\frac{\partial}{\partial J}\right); ~~\frac{\partial}{\partial g^{yy}}=-\frac{1}{2}\left(\frac{\partial}{\partial \tau_2}+\frac{\partial}{\partial J}\right)\nonumber\\ 
&&\frac{\partial}{\partial g^{xy}}=\frac{\partial}{\partial g^{yx}}=-\frac{1}{2}\frac{\partial}{\partial \tau_1}.
\end{eqnarray}
Notice that the derivative with respect to $g^{xy}$ is taken at constant $g^{yx}$ and viceversa~\cite{note2}. 

Inserting Eq.~(\ref{derivatives}) into Eq.~(\ref{key}), and noting that the wave function  does not depend on $J$ ($J$ enters the homogeneous Hamiltonian as a global scale factor),  we get the following representation for $\Lambda_0$
\begin{eqnarray}\label{Lambda0}
\Lambda_0 =2 \times \Im m\left. \left\langle \frac{\partial \psi_0}{\partial \tau_1}\right\vert   \frac{\partial \psi_0}{\partial \tau_2}\right\rangle,\end{eqnarray}
where the derivatives are calculated at $\tau_1=0$, $\tau_2=1$ and $J=1$.

The calculation of $\Lambda_0$ is dramatically simplified if we assume that the wave function lies within the lowest Landau level.  First of all, let us define ``Landau levels".  It is easy to see that the eigenfunctions of the kinetic energy operator~(\ref{TN}) with a homogeneous metric and the stated boundary conditions have the form (up to a normalization constant)
\begin{equation}
\psi_{lk}(x,y) = e^{2 \pi i k x}e^{i \pi \tau N_L \tilde y^2}H_l(\sqrt{2 \pi \tau_2 N_L} \tilde y)
\end{equation}
where $\tilde y = y+k/N_L$,  $k$ is an integer, and $H_l$ is the l-th Hermite polynomial. (The eigenvalue is $(l+1/2)\omega_c/J$).  The (degenerate) states of the lowest Landau level are the ones with $l=0$ and $k$ ranging from $0$ to $N_L-1$. 
From this, we immediately see that any $N$-electron wave function that lies entirely in the lowest Landau level  must have the form
\begin{equation} \label{PsiLLL}
\psi (\rv_1,...,\rv_N) = A(\tau_1,\tau_2) f(\xi_1,...,\xi_N)\prod_{i=1}^N e^{i \pi \tau N_L y_i^2}
\end{equation} 
where $f$ is an {\it analytic} function of the variables
$
\xi_i \equiv x_i+\tau y_i,
$
and therefore also of $\tau = \tau_1+i\tau_2$.   $A(\tau_1,\tau_2)$ is the normalization constant.  Unlike $f$, $A$ depends separately on $\tau_1$ and $\tau_2$, i.e. it is a non-analytic function of $\tau$. [We note that the form of Eq.~(\ref{PsiLLL}) is, in fact, independent of boundary conditions.] Hence in general the $\tau$-dependence of any LLL many-body wave function can be represented as $\psi(\tau_1,\tau_2) = A(\tau_1,\tau_2)\Phi(\tau)$, 
where $\Phi(\tau)$ is a holomorphic function. 
  Making use of this fact it is possible to show that Eq.~(\ref{Lambda0}) simplifies as follows \cite{Levay1995}
\begin{eqnarray}\label{Lambda01}
\Lambda_0 =-\frac{\hbar}{L^2}\left(\frac{\partial^2}{\partial \tau_1^2}+\frac{\partial^2}{\partial \tau_2^2}\right)\ln A(\tau_1,\tau_2)~.
\end{eqnarray}
(Here we have reinstated the physical units, multiplying the result of the dimensionless calculation by the factor $\hbar/L^2$, which had been previously set to $1$).

\begin{figure}
\begin{center}
\includegraphics[width=0.8\linewidth]{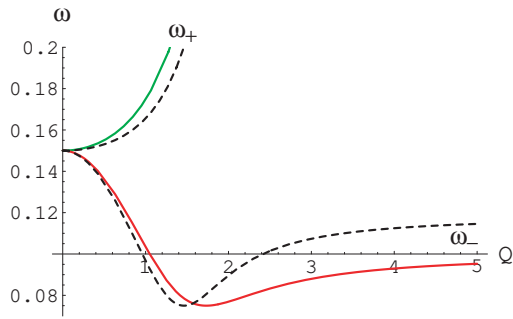}
\caption{(Color online) Dispersion of the collective modes at $\nu=1/3$ obtained from the solution of Eq.~(\ref{eom1}) for (i)  the value of $\Lambda_0 = 1.4 \hbar/(4n)$ used  in Ref.~\onlinecite{TokVigPRL2007} (dashed lines) and (ii)  the exact value from Eq.~(\ref{Lambda02}) (solid lines).   All other parameters are the same as in Ref.~\onlinecite{TokVigPRL2007}. Here $Q=q \ell$ and $\omega$ is in units of $ e^2/(\hbar \epsilon_b\ell)$.  \label{fig:one}}
\end{center}
\end{figure}

The problem is finally reduced to calculating derivatives of the normalization constant.  A second major simplification occurs in the thermodynamic limit, i.e. when the number $N$ of electrons tends to infinity in such a way that $N/N_L = \nu$ remains constant.  Because $N_L$ tends to infinity it is evident that only values of $y \sim 1/N_L \sim 0$ contribute to the normalization integral, and therefore the dependence of the analytic factor $f$ on $\tau$ (through $\xi=x+\tau y$) effectively disappears.   Then we immediately see that the $\tau$-dependence of the normalization constant is of the form $A(\tau_1,\tau_2) \propto  \tau_2^{N/4}$,
and use of Eq.~(\ref{Lambda01}) yields
\begin{equation}\label{Lambda02}
\Lambda_0 = \frac{\hbar n}{4}
\end{equation}
where $n=N/L^2$ is the areal density of electrons.  We emphasize that this result does not depend on the detailed form of the wave function.

Let us now apply the above universal result to the dispersion of collective modes in incompressible liquid states. In Ref.~\onlinecite{TokVigPRL2007} we developed a theory of the elastic constants in the incompressible quantum Hall liquid, based on the premise that  the the stress-stress response function is dominated by a single pole at the frequency $\Delta$ of the $q=0$ gap. This assumption uniquely fixes the frequency dependence of the ordinary shear modulus $\mu(\omega)$ and the Lorentz shear modulus $\Lambda(\omega)$
\begin{equation}
\label{mu(omega)}
\mu(\omega) = \frac{\mu_{\infty}\omega^2}{\omega^2 - \Delta^2},\qquad 
\Lambda(\omega) = \frac{\Lambda_0\Delta^2}{\omega^2 - \Delta^2},	
\end{equation}
while the bulk modulus $K$ remains frequency independent for any LLL state. 
The excitation gap $\Delta$, the bulk modulus $K$ together with $\mu_{\infty}$ and $\Lambda_0$ became the phenomenological parameters of the theory ($\mu^\infty$ and $\Delta$ however could be determined self-consistently from sum rule arguments provided the microscopic ground state wave function is given \cite{TokVigPRL2007}).

Inserting the above $\mu(\omega)$, $\Lambda(\omega)$, and $K$ into Eq.~(\ref{Qh1}) and this into Eqs.~(\ref{Pij}) and (\ref{eom1}), and solving the resulting dispersion equation we get two collective modes with frequencies $\omega_{\pm}(q)$
\begin{eqnarray}
	\label{omega(q)}
&&	\frac{\omega_{\pm}^2}{\Delta^2} = 1 + \bar\mu_{\infty}(\bar\mu_{\infty}+\bar{K})\frac{Q^4}{2} 
    - \bar\Lambda_0 Q^2 \\\nonumber
	&\pm& \sqrt{\Big[\bar\mu_{\infty}(\bar\mu_{\infty}+\bar{K})\frac{Q^4}{2} - \bar\Lambda_0Q^2\Big]^2 +
	(\bar\mu_{\infty}^2 - \bar\Lambda_0^2)Q^4}
\end{eqnarray}	
where $Q=q\ell$, $\bar\mu_{\infty}=\frac{\mu_{\infty}}{\hbar n\Delta}$, $\bar{K}=\frac{\bar{K}}{\hbar n\Delta}$, and $\bar\Lambda_{0}=\frac{\Lambda_{0}}{\hbar n}$ are the dimensionless wave vector and elastic moduli.

In the special case of a Laughlin wave function with filling factor $\nu=1/M$, where $M$ is an odd integer larger than $1$, we have shown in Ref.~\onlinecite{TokVigPRL2007} that $\frac{\mu_\infty}{\Delta} = \frac{\hbar n (M-1)}{4}$.
Combining this with our new result of Eq.~(\ref{Lambda02}) we get $\frac{\Lambda_0 \Delta}{\mu_\infty} = \frac{1}{M-1}$.
Hence the only phenomenological parameter that is left at the current stage of the theory is the bulk modulus $K$. Importantly, $K$ is irrelevant for small $q$. Therefore at long wavelength our dispersion relations (\ref{omega(q)}) for the Laughlin states are completely parameter-free.

Fig.~\ref{fig:one} shows the dispersion of the collective modes calculated at $\nu=1/3$ with the exact value of $\Lambda_0$ in comparison with the somewhat larger value of  $1.4 \frac{\hbar n}{4}$ used in Ref.~\onlinecite{TokVigPRL2007}  (the bulk modulus remains the same).  It is clear that the overall behavior of the curve is still quite satisfactory.  The reduced values of $\Lambda_0$ flattens the dispersion of the collective modes at large $q$ giving a shallower roton minimum.  The changes in the dispersion at small $q$ are quite small.

In summary, we have presented an exact calculation of the low-frequency elasticity modulus $\Lambda_0$ of interacting electrons in the lowest Landau level.  The fact that the calculation can be done in universal form is a consequence of two facts:  (i) The existence of a geometric representation of $\Lambda_0$, and (ii) The analyticity of lowest Landau level wave functions with respect to the complex deformation parameter $\tau$, which describes a variation of the metrics.  We believe that this general result will be useful to further elucidate the dynamics of collective states (not necessarily uniform) in the lowest Landau level.

Support from NSF Grant No. DMR-0313681 is gratefully acknowledged. 


\begin{thebibliography}{99}
\bibitem{note2}
The factor $1/2$ in the last of Eqs.~(\ref{derivatives}) originates from the fact that a small increment in $\tau_1$ changes both $g^{xy}$ and $g^{yx}$ by equal amounts, whereas in Eq. (12) only $g^{xy}$ is changed, $g^{yx}$ remaining constant.
\end{thebibliography}

\end{document}